\documentclass[epjST]{svjour}
\usepackage{graphicx}
\usepackage{amssymb}
\usepackage{bm}

\begin{document}
\title{Dynamics of ions in the selectivity filter of the KcsA channel}
\subtitle{Towards a coupled Brownian particle description}
\author{S. M. Cosseddu\inst{1,2}\fnmsep\thanks{\email{s.cossedu@warwick.ac.uk}} \and I. A. Khovanov\inst{1,2} \and M. P. Allen\inst{3} \and P. M. Rodger\inst{1,4}  \and D. G. Luchinsky\inst{5,6}  \and P. V. E. McClintock\inst{6} }
\institute{Centre for Scientific Computing, University of Warwick, Coventry, CV4 7AL, UK \and School of Engineering,  University of Warwick, Coventry, CV4 7AL, UK  \and Department of Physics, University of Warwick, Coventry CV4 7AL, UK \and Department of Chemistry, University of Warwick, Coventry CV4 7AL, UK \and Mission Critical Technologies Inc., 2041 Rosecrans Ave. Suite 225 El Segundo, CA  90245, USA \and Physics Department, Lancaster University, LA1 4YB, Lancaster, UK  }
\abstract{
The statistical and dynamical properties of ions in the selectivity filter of the KcsA ion channel are considered on the basis of molecular dynamics (MD) simulations of the KcsA protein embedded in a lipid membrane surrounded by an ionic solution.  A new approach to the derivation of a Brownian dynamics (BD) model of ion permeation through the filter is discussed, based on unbiased MD simulations. It is shown that depending on additional assumptions, ion's dynamics can be described either by under-damped Langevin equation with constant damping and white noise or by Langevin equation with a fractional memory kernel.  A comparison of the potential of the mean force derived from unbiased MD simulations with the potential produced by the umbrella sampling method demonstrates significant differences in these potentials. The origin of these differences is an open question that requires further clarifications.
} 
\maketitle
\section{Introduction}
\label{intro}

Ion channels are transmembrane proteins that are able to catalyze ion flow across the membrane, with high selectivity and efficiency, according to the electrochemical gradient~\cite{hille_ion_2001}. They are involved in a variety of biological mechanisms. For example, channels are responsible for the regulation of osmotic pressure and cell volume, as well as for the membrane potential and electrical activity of the cell~\cite{hille_ion_2001}. A famous example of an ion channel is the KscA potassium (K$^+$) channel from Streptomyces lividans bacteria, which has an amino acid sequence closely similar to that of vertebrate and invertebrate voltage-dependent potassium channels~\cite{doyle_structure_1998}. The detailed molecular structure of KcsA has been known since 1998~\cite{doyle_structure_1998}. Since then, it has been actively studied through the use of molecular dynamics (MD) simulations for verifying and testing a variety of hypotheses related to the conductivity and selectivity of the channel. Despite the fact that the structure of KcsA is known, however, a purely {\em mechanistic} study of KcsA is prohibited by the extremely high dimension of the corresponding model (see below for details) and by the limitations of even modern computational resources, as well as by the inherent limitations of the MD approach~\cite{Eisenberg_2010}. Consequently, MD is often used as a tool for deriving coarse-grained models, which in turn should provide a link to experimental observations. One such approach is the derivation of a Brownian dynamics (BD) model~\cite{Eisenberg_2010}, which can be analysed either analytically or numerically because it has just a few {\em effective} degrees of freedom.

The MD level of description is based on the Hamiltonian formalism, where each atom of the biological system obeys the deterministic equations of motion:
\begin{eqnarray}
\label{eq:hmd}
\frac{d{\bf r}_i}{dt}=\frac{\partial H }{\partial {\bf p}_i}\equiv \frac{{\bf p}_i}{m_i}, \ \ \ \frac{d {\bf p}_i}{dt}=-\frac{\partial H }{\partial {\bf r}_i} \equiv -\nabla_{{\bf r}_i} U, \ \ \ H  = \sum_{k=1}^{N} \frac{{\bf p}_k \cdot {\bf p}_k}{2 m_k} + U( {\bf r}_1, {\bf r}_2, \ldots {\bf r}_N) \:.
\end{eqnarray}
Here $m_i$ is the mass of the $i$th atom, ${\bf r}_i=(x_i, y_i, z_i)$ and ${\bf p}_i =(p_{ix}, p_{iy}, p_{iz})$ are respectively the coordinates and momenta of the $i$-atom, $U$ is the potential energy function describing inter-atom interactions, and $H$ is the Hamiltonian of the $N$-atom system. The magnitude of $N$ reflects the number of atoms involved, and it is of the order of $10^6$ or $10^7$. The quantity  $-\nabla_{{\bf r}_i}U$ specifies the forces acting on each atom; the forces are taken from a library of force fields, which are calculated on the basis of quantum mechanics and then corrected to fit the experimental results~\cite{mackerell_jr._empirical_2004}. Typically, the trajectories generated by MD simulations are used to sample in one of the thermodynamic ensembles~\cite{mcquarrie_statistical_2000} in order to estimate an order parameter for describing  macroscopic properties. The task of analysing the permeation of ion(s) through the channel is different, and interest centres on the properties of a selected part of the system (the permeating ion). In this case, a BD model of the ionic dynamics can be derived from MD trajectories~\cite{kneller_scaling_2008}. It is assumed~\cite{Adelman_1980} that the MD trajectory $({\bf r}_i, {\bf v}_i)$ of the ion can be described by a generalized Langevin Equation (GLE) of the form:
\begin{eqnarray}
\label{eq:gle}
m_i\dot{\bf v}_i(t) = -\frac{\partial V({\bf r}_i)}{\partial {\bf r}_i} -\int_0^t {\bf M}(t-\tau){\bf v}_i(\tau) d\tau+{\bf R}(t),
\end{eqnarray}
where ${\bf v}_i$ is the ion's velocity (${\bf v}_i={\bf p}_i/m_i$), $V({\bf r}_i)$ is the so-called potential of mean force (PMF), and ${\bf M}(t)$ is an appropriate memory function, connected to the random force ${\bf R}(t)$  acting on the ion via the second fluctuation-dissipation theorem:
\begin{eqnarray}
\label{eq:sfdt}
{\bf M}(t)= \frac{1}{k_B T}\langle {\bf R}(0) {\bf R}(t) \rangle.
\end{eqnarray}
Here $k_B$ and $T$ are respectively Boltzmann's constant and the temperature in the MD simulation.

Thus, an MD model is a Hamiltonian, non-linear, high-dimensional, dynamical system, which typically shows  multi-scale behaviour in both space and time. The high-dimension prevents the application of standard dynamical systems approaches such as stability and bifurcation analyses. To reduce the dimension, therefore, ergodicity and mixing are assumed in such systems. These assumptions allow us to consider each atom as a particle moving in some potential under the action of a stochastic source; that is, it allows us to proceed to a BD description. In this way, instead of studying a high-dimensional non-linear Hamiltonian system, one can consider the properties of a low-dimensional dissipative stochastic non-linear system.

Questions related to ergodicity in dynamical systems, and stochastic effects in non-linear systems, are topics to which Prof. Vadim Anishchenko has made numerous seminal contributions. On the occasion of his 70th birthday, the authors extend their warmest congratulations and best wishes for his continued fruitful and enjoyable research in these directions.

The key task of the MD to BD transition in the level of description amounts to identifying the potential $V({\bf r}_i)$, the memory function ${\bf M}(t)$ and the fluctuation force ${\bf R}(t)$.
In general these identifications are very complicated but, in practice they can be simplified by a number of assumptions, the validity of which can in principle be tested separately.
A typical assumption is that ionic motion corresponds to overdamped Markovian diffusion~\cite{berneche_molecular_2000}, resulting in the following simplified equation~\cite{Risken:96}
\begin{eqnarray}
\label{eq:overd}
\dot{\bf r}_i = -\frac{1}{m_i \gamma({\bf r}_i)}\frac{\partial V({\bf r}_i)}{\partial {\bf r}_i} +\sqrt{\frac{2 k_B T }{m_i  \gamma({\bf r}_i)}}\bm{\xi}(t),
\end{eqnarray}
where $\gamma({\bf r}_i)$ specifies the damping and  $\bm{\xi}(t)=(\xi_x(t), \xi_y(t), \xi_z(t))$ is a vector of Gaussian white noises. Values of $\gamma$ can be calculated if the diffusion constant $D$ is known via the relation $D=k_BT/\gamma$, where in turn $D$ can be taken as being equal to the bulk diffusion (known from experiment) or it can be estimated using coordinate and velocity time-series (see \cite{Mamonov_2006} and below for details). Note that there are formal analytical approaches \cite{Zwanzig:75,Goodyear:96} allowing one to derive the stochastic equation (\ref{eq:gle}) starting from Hamiltonian equation (\ref{eq:hmd}). However their use is very limited because they explicitly or implicitly assume \cite{Goodyear:96} that a ``particle'' of interest is weakly bi-linearly coupled to a bath of harmonic oscillators. In practice, as in the case considered of an ion inside the selectivity filter, the coupling is strong and non-linear.  Therefore techniques for calculating the PMF $V({\bf r}_i)$ on the basis of the often trajectories of MD simulations~\cite{piccinini_biased_2008} are often used. These techniques are based on the introduction of a known additional deterministic force in the Hamiltonian $H$. This additional force facilitates an activation event  and moves the system from a given initial state to some another state. The PMF is calculated using a combination of MD simulation trajectories and values of the additional force.   The most often applied technique is the so-called Umbrella Sampling (US) method, used to consider ion channel properties and to confirm the key hypothesis of  ``knock-on'' permeation in the KcsA channel~\cite{berneche_energetics_2001}. The US method applies a harmonic bias potential~\cite{laio_metadynamics:_2008} and, for a set of values of the biased potential, the coordinate distributions $p({\bf r}_i)$ are built by MD simulations; then the unbiased potential, that is the PMF, is recovered by specific approaches, usually the WHAM method~\cite{kumar_multidimensional_1995}. Note that coordinate distributions $p({\bf r}_i)$ are applied in all other techniques for estimating the PMF, for example in the metadynamics approach~\cite{laio_metadynamics:_2008}. Thus several different assumptions are employed to proceed from the MD model (\ref{eq:hmd}) to the higher-level BD model (\ref{eq:overd}), and the distribution $p({\bf r_i})$ lies at the heart of the MD to BD transition.

In this manuscript we consider ionic dynamics in the selectivity filter of the KcsA channel using MD simulations, and we discuss an approach to the derivation of the corresponding BD model. First, we present some details of the MD simulations and of the KcsA channel, as well as of the PMF calculation via the US method. Secondly, we analyse the statistical and dynamic properties of ion motion and verify the applicability of the overdamped BD model (\ref{eq:overd}) for describing ion motion. Thirdly, we introduce an approach to derive the parameters of the BD model using the steady-state motion of the ion. Finally, the PMFs obtained by this approach and by the US method are compared.

\section{Methods and system}
\label{methods}

\subsection{The KcsA channel and its selectivity filter}

\begin{figure}[h!]
\includegraphics[width=8cm]{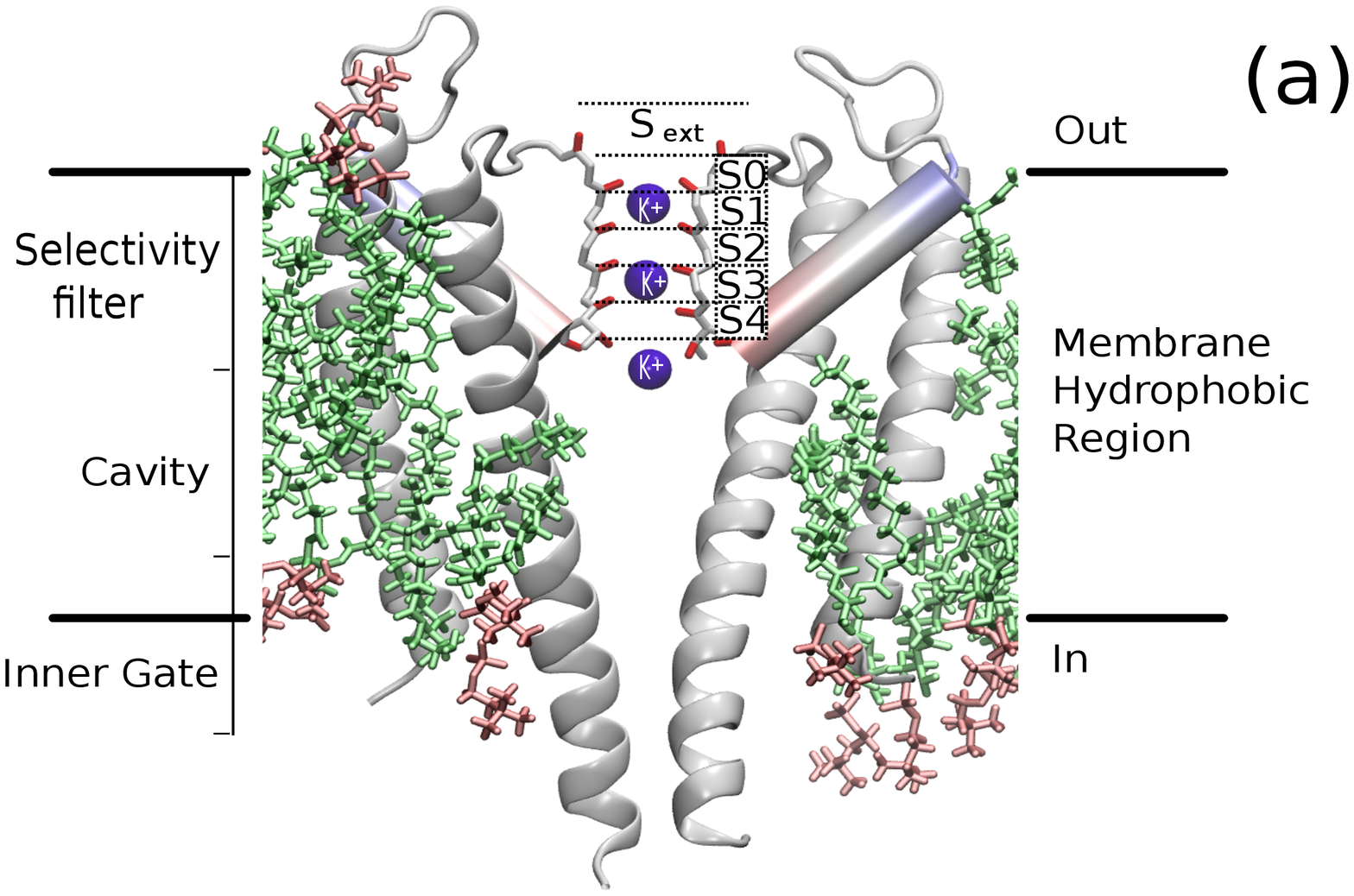}~~~~~~\includegraphics[width=3.5cm]{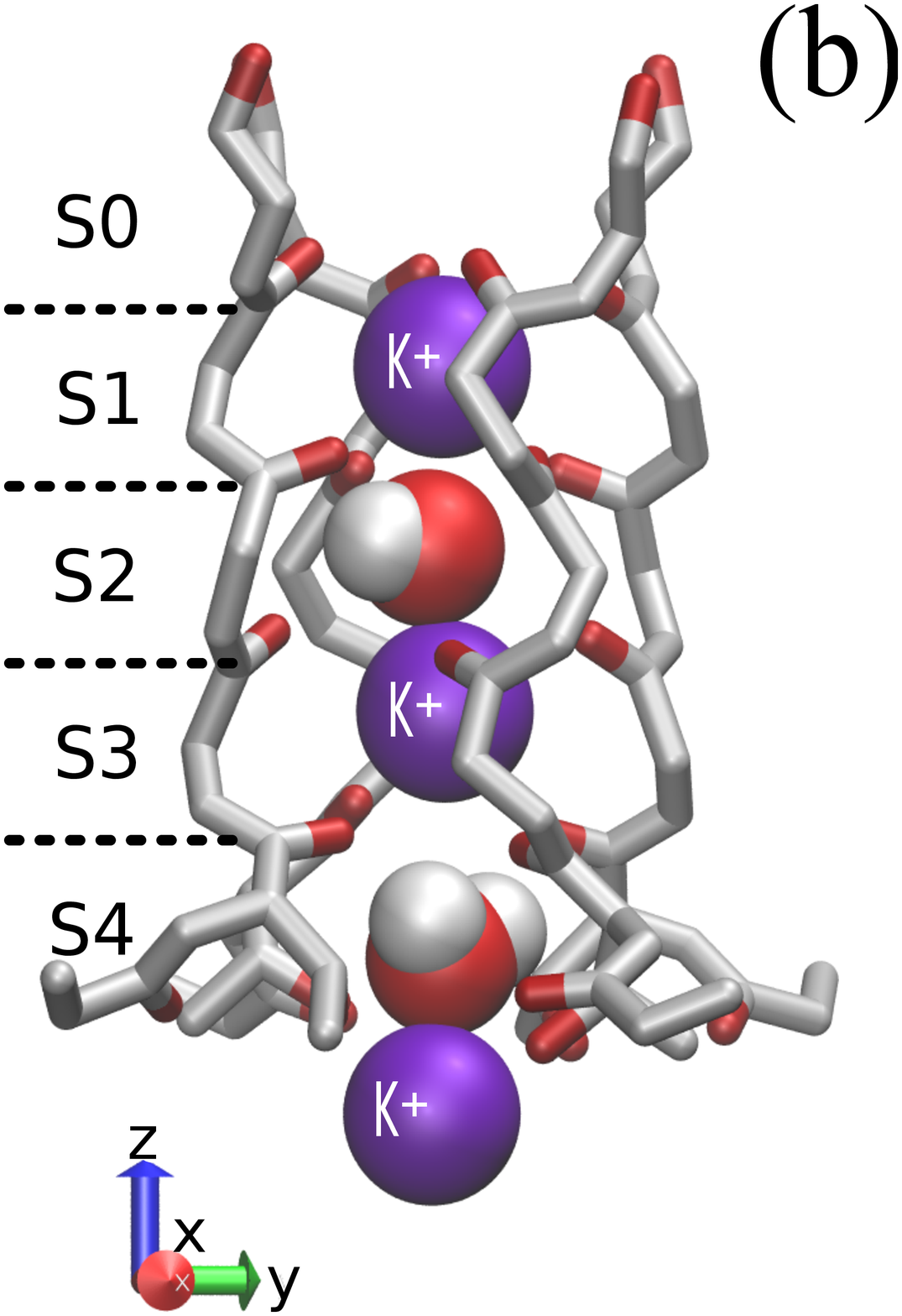}
\caption{(Colour online) (a) The protein KcsA embedded in the lipid membrane, showing only two of its 4 subunits. (b) The configuration C1 of ions and water molecules in the selectivity filter. The K$^+$ ions are shown by (violet) spheres, denoted K$^+$,  and oxygen atoms  are shown by dark (red) colour. In each case, the possible ionic binding sites are labelled S0--S4.}
\label{fig1}       
\end{figure}

The simple bacterial channel KcsA from Streptomyces lividans~\cite{hille_ion_2001,doyle_structure_1998} resembles human K$^+$ channels in terms of ion permeation and selectivity.  It is an integral membrane tetrameric protein in which each subunit is formed by two transmembrane helices  connected by a P-loop. The P-loop is made up of a short polarised helix (P-helix) and an amino acid \emph{signature sequence}, TVGYG, that represents a motif conserved across all K$^+$ channels ~\cite{hille_ion_2001,doyle_structure_1998}.  The channel structure can be divided into three main regions with different functions (Fig.~\ref{fig1}(a)): the selectivity filter or mouth, formed by the TVGYG sequences of the four subunits, which is considered to be responsible for selectivity between different cationic species; a wider inner cavity filled with water where the ions are hydrated to provide an environment that is energetically favourable to cations in the hydrophobic region of the membrane; and the intracellular gate, associated with pH-dependent rigid-body movements of the transmembrane helices able to control the permeation pathway.

The filter has four internal binding sites for ions (named S1 -- S4) and two external sites (S0 and S$_{\rm ext}$)~\cite{doyle_structure_1998}. Sites S0 -- S4 consist of the strongly dipolar TVGYG groups: the carbonyl groups of the backbone of V76 G77 Y78 G79 that points into the pore, and the hydroxyl group of the T75 side chain. Each permeating ion interacts with the oxygen atoms from the 8 dipolar groups in a square anti-prism geometry~\cite{zhou_chemistry_2001}, resulting in single file permeation. The positions of the ions inside the sites are stable (metastable), and their possible locations inside the filter are denoted by S1 -- S4 in Fig.~\ref{fig1}.

For the MD simulations, we chose to analyse the ion dynamics in particular configurations corresponding to the three sites S2, S1 and S0. The PMF was obtained for the same three positions, using the US method. Two configurations were selected for MD simulations. One of these is C1, shown in Fig.~\ref{fig1}(b): there are two ions in the cavity, a water molecule in S4, an ion in S3, a water molecule in S2 and an ion in S1. The other configuration, referred to as C2 (not shown) corresponds to: an ion in S4, a water molecule in S3, an ion in S2, a water molecule in S1, and an ion in S0.

\subsection{MD simulations}

The MD simulations were performed using NAMD 2.8 \cite{phillips_scalable_2005} in the $NpT$ ensemble with pressure $p=$1.01~bar, temperature $T=$310~K, and $N=89390$ atoms. A multiple timestep algorithm is used \cite{tuckerman_reversible_1992} with: an integration step of 1~fs; nonbonded interactions calculated every 2~fs; and electrostatic forces every 4~fs, using a smooth particle-mesh Ewald (SPME) method \cite{essmann_smooth_1995}. The CHARMM22 force field (FF) was used for the protein, with a modification in the Lennard-Jones term to represent the interaction between K$^+$ and the carbonyl oxygens of the protein, CHARMM36 for the lipids and TIP3P for water \cite{mackerell_all-atom_1995,klauda_update_2010,Noskov:04}.

A set of MD simulations was performed. For each integration step of 1fs=0.001ps, the coordinates ${\bf r}=(x,y,z) $ in units of \AA~ ($10^{-10}$m) ~and velocities ${\bf v}=(v_x, v_y, v_z) $ in \AA/ps ($10^{+2}$ m/s) of different atoms were stored. These included: ions and water molecules inside the selectivity filter,  ions in the cavity and ions in the bulk, as well as the oxygen atoms of the active sites. A typical realization consisted of $10^6$ time steps.  During the simulations, the lipids of the membrane are not constrained and so the simulated piece of membrane slowly moves together with the channel protein. To compensate for this slow motion, all coordinates were therefore calculated relative to the centre of mass of the selectivity filter. Note that, as a result, slow time scales longer than about $10^5$ ps were automatically removed from consideration.

\subsection{PMF calculations}
\label{pmf}

The choice of  parameters for the US method followed the paper by Piccinini \emph{et al.} \cite{piccinini_biased_2008}: a time step of 1 fs, force constant 20 kcal/mol \AA{}$^2$; $\Delta z= 0.5$ \AA{} between centres of the biasing potential; 14 biased steps of length 515 ps; 15 ps from every window considered as the initialization time needed to reach the new centre; and 250 ps considered as an equilibration time. The coordinate $z$ corresponding to a biased collective variable was stored at every time step. The PMF was obtained by the WHAM method~\cite{kumar_multidimensional_1995} using a histogram of 310 bins spanning from 0--8\,\AA{}, i.e.\ between positions S2 and S0.

\section{Analysis of equilibrium dynamics}
\label{equilib}

The development of a BD model tacitly assumes that the atomic motion is stochastic, i.e.\ that the trajectories $({\bf r}, {\bf v})$ of the MD simulations are realizations of a stochastic process. The latter are characterized~\cite{Bendat_2011} by the probability density functions (distributions) $p({\bf r})$ and $p({\bf v})$, as well as by the power spectrum $S(f)$ and/or the auto-correlation coefficient $\rho (\tau)$ of  the coordinates' or velocities' projections onto one of the Cartesian axes $x$, $y$ or $z$. Statistical mechanics predicts  that, for an ensemble of atoms in equilibrium, the distribution of coordinates and velocities of the atom for any projection $r \equiv (x, y, z)$ takes the following form \cite{Allen:89}:
\begin{eqnarray}
\label{eq:boltz}
 p(r,v_r)= \frac{1}{\mathcal{N}} \exp\left(-\frac{m}{2 k_B T_r} v_r^2\right) \exp\left(-\frac{V(r) }{k_B T_r} \right),
\end{eqnarray}
where $\mathcal{N}^{-1}$ is the normalization factor. $V(r)$ is the potential for a given atom in the direction $r$ and corresponds to the PMF. This prediction enables the PMF to be calculated, for example by the US method, via the numerically estimated distribution $p(r)$ by use of the following expression:
\begin{eqnarray}
\label{eq:PMFc}
V(r) = C_r - k_B T_r \ln[p(r)],
\end{eqnarray}
where $C_r$ is a constant related to the distribution's normalization and is ignored during an estimation. The value of $T_r$ is usually considered equal to $T$, the system's temperature as specified in the MD simulations; it can be estimated via the variance $(k_B T_r /m )$ of the distribution $p(v_r)$ which, in turn,  is predicted to be Gaussian. This last fact can be verified.

The resultant time-scales of the atomic dynamics can be revealed via $S(f)$ and/or $\rho (\tau)$. Although there are no analytic expressions for $S(f)$ and $\rho (\tau)$ for arbitrary forms of PMF $V(r)$, a huge  volume of results is nonetheless available for different stochastic nonlinear equations similar to Eq.\ (\ref{eq:overd}). They place  limitations on the possible shapes of the power spectrum and the auto-correlation function.  Thus, combining estimated distributions with power spectra (or auto-correlation function)  it is possible to verify the validity of the use of the overdamped Langevin dynamics for the ion in the selectivity filter. We emphasise that this latter assumption is widely used \cite{berneche_energetics_2001} for analysing the properties of ions in the KcsA channel.

\begin{figure}[h!]
\includegraphics[width=6.5cm]{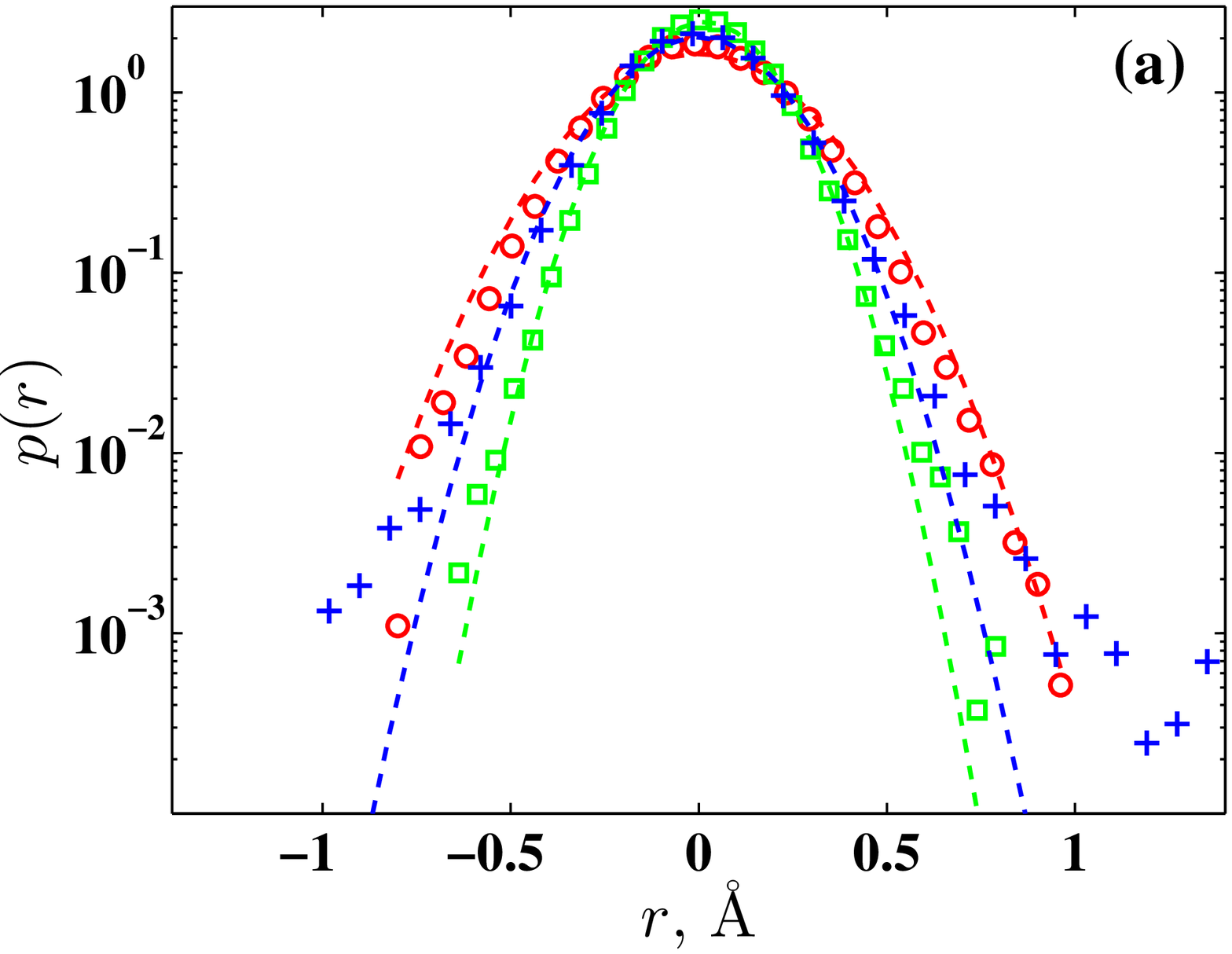}~~\includegraphics[width=6.5cm]{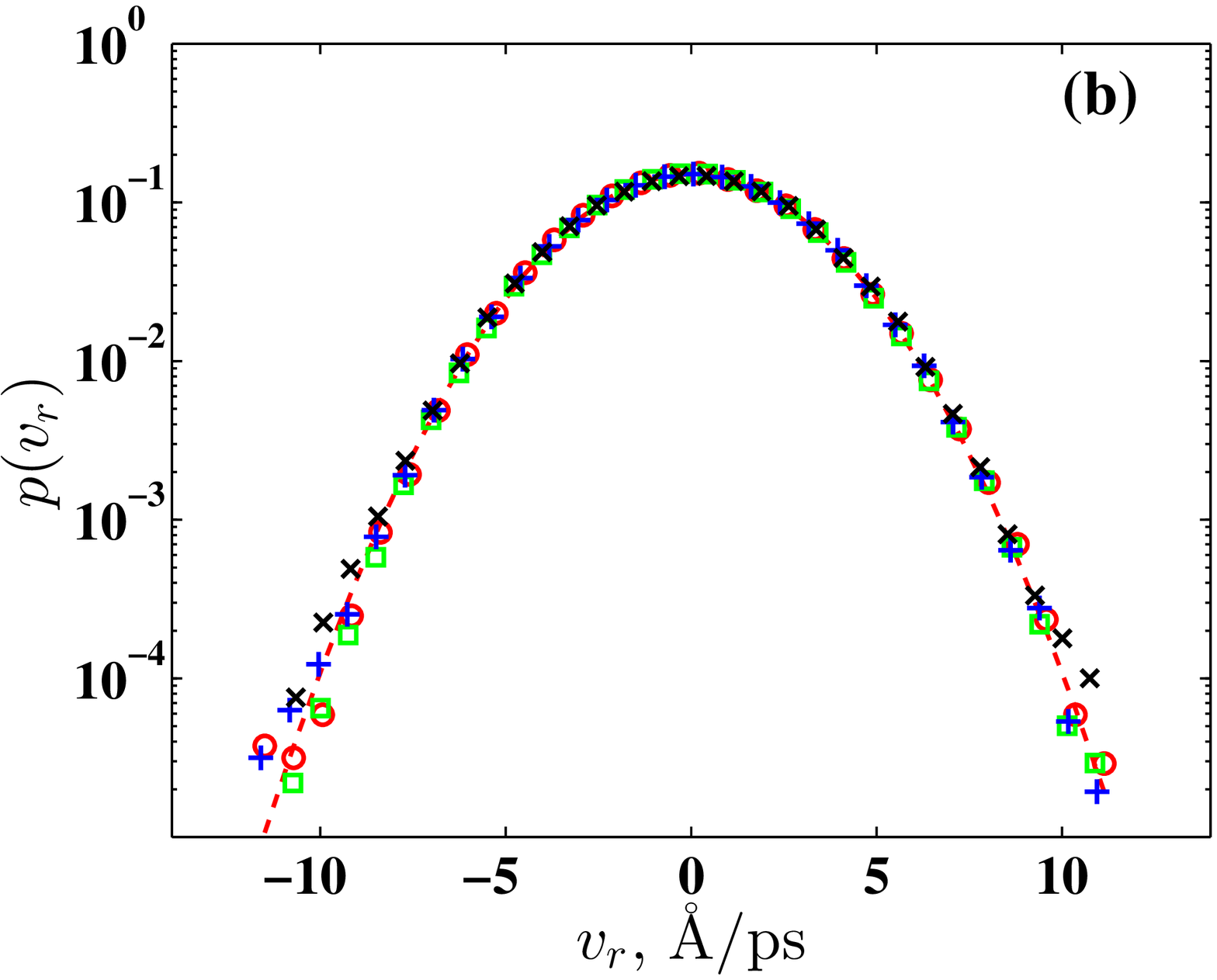}
\caption{(Colour online) Distributions (log scale) of (a) coordinate $p(r)$  and (b) velocity $p(v_r)$  of ions for different locations are shown by markers: $\circ$ (red colour) corresponds to the coordinate $x$ and velocity $v_x$ of an ion  in the site S1, $\square$ (green colour) corresponds to the coordinate $y$ and velocity $v_y$ of an ion  in the site S2, $+$ (blue colour) corresponds to the coordinate $z$ and velocity $v_z$ of an ion  in the site S0, and $\times$ (black colour) corresponds to the velocity $v_z$ of an ion in the bulk (the coordinate's distribution is not shown in this case since it is time-dependent). The coordinate distribution $p(r)$ are centred with respect to the mean value, that is $\langle r \rangle = 0$. Dashed lines correspond to fits of Gaussian distributions with mean values and the variances estimated from the corresponding time-series.}
\label{fig2}       
\end{figure}

Distributions of the coordinate and velocity of a K$^+$ ion in different locations are shown in Fig.~\ref{fig2}. The velocity has the same distribution, regardless of the position of the ion: almost perfectly Gaussian, thus confirming the correctness of the first term in the expression (\ref{eq:boltz}). So the value of temperature $T_r$ can properly be estimated on the basis of $p(v_r)$. An analysis of distributions in terms of the coordinates $x$, $y$ and $z$ for an ion at the sites S0, S1 and S2 shows that the distributions are similar, and close to Gaussian for coordinates $x$ and $y$, but that they differ significantly from Gaussian for the $z$ coordinate, as illustrated in Fig.~\ref{fig2}(a). Such a picture is not unexpected given that the channel is of cylindrical symmetry with $z$ as its main symmetry axis. The permeation process occurs in the $z$-direction through a sequence of jumps by the ion between different sites. Hence the potential $V(z)$ is multistable, and thus non-parabolic. In fact, the shape of $V(z)$ is one of the keys to an adequate description of ion permeation. Analyses of the distributions of $p(z)$ for different steady state positions demonstrate that the potential is close to being harmonic (parabolic) in the vicinity of each stable/metastable state, but that its shape deviates from parabolicity at  large displacements from that state. The fact that, in contrast, the distributions $p(x)$ and $p(y)$ are close to Gaussian, means that the PMF in $(x,y)$ is of a parabolic shape and consequently that the systems (\ref{eq:overd}) and (\ref{eq:gle}) are linear.

\begin{figure}[h!]
\includegraphics[width=6.5cm]{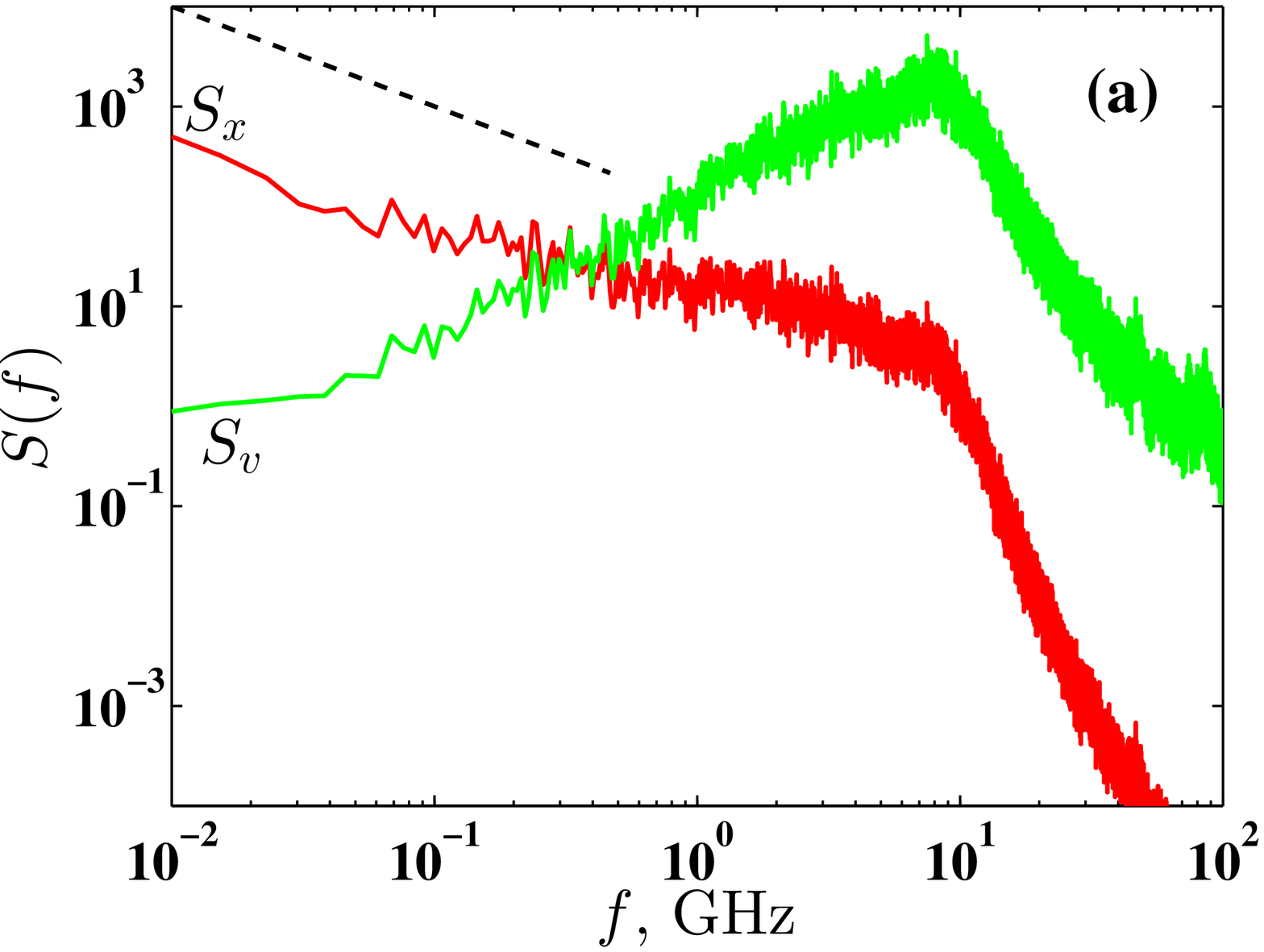}~~\includegraphics[width=6.5cm]{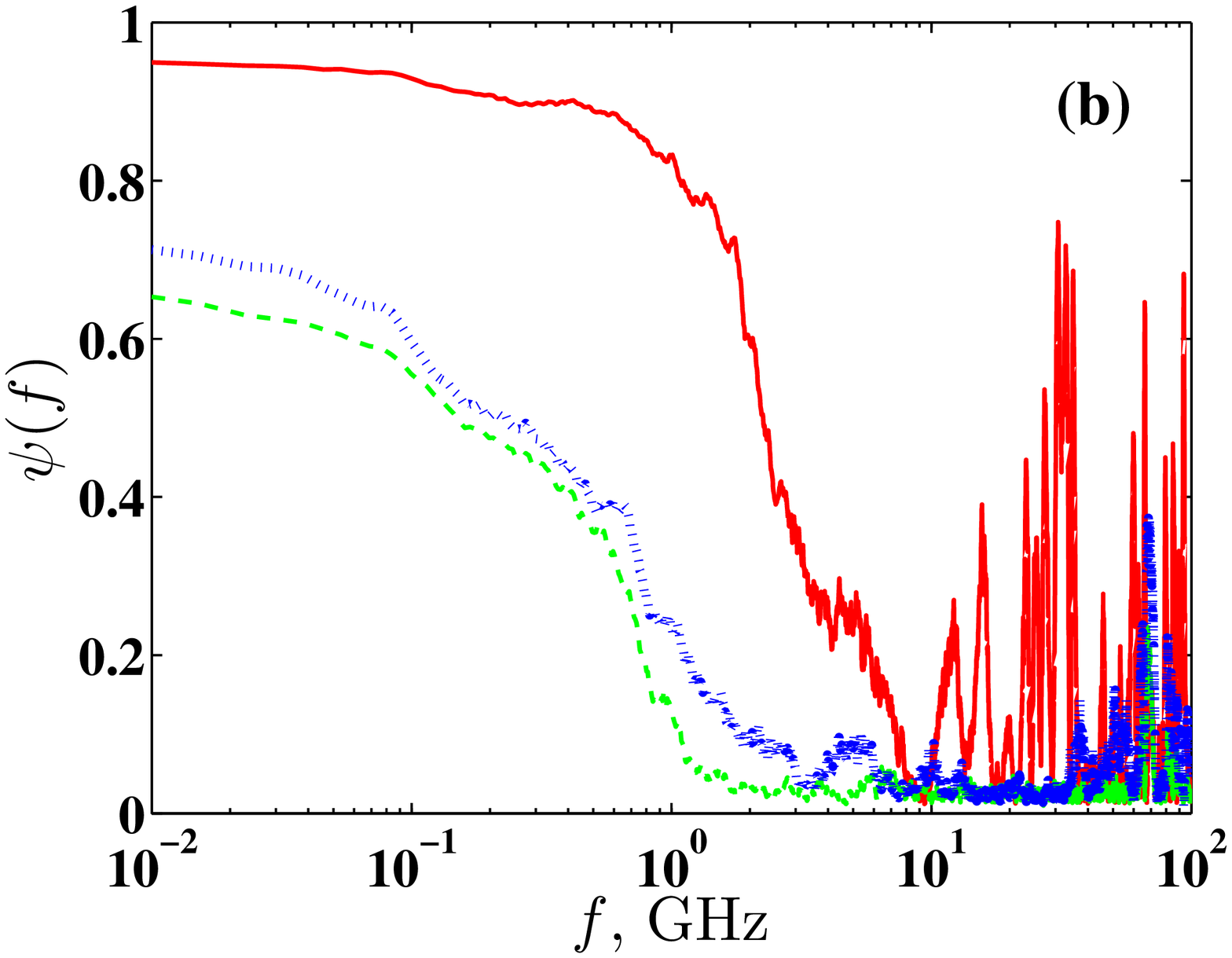}
\caption{(Colour online) (a) The power spectra $S_x$ of coordinate $x$ (red colour) and $S_v$ velocity $v_x$ (green colour) of the ion in the site S1. The dashed line indicates $1/f$ scaling. Both axis scales are logarithmic. (b) The spectral coherence $\psi(f)$ between the $z$ coordinate of an oxygen atom at the bottom of the site S1 and the $z$ coordinates of three different atoms: the solid (red) line corresponds to an oxygen atom located on the same sub-unit at the top of the site S1; the dashed (green) line corresponds to an oxygen atom located on a different sub-unit at the bottom of the site S1; and the dotted (blue) line corresponds to an ion located in the site S1. The $x$-axis is logarithmic.}
\label{fig3}       
\end{figure}

\begin{figure}[h!]
\includegraphics[width=6.5cm]{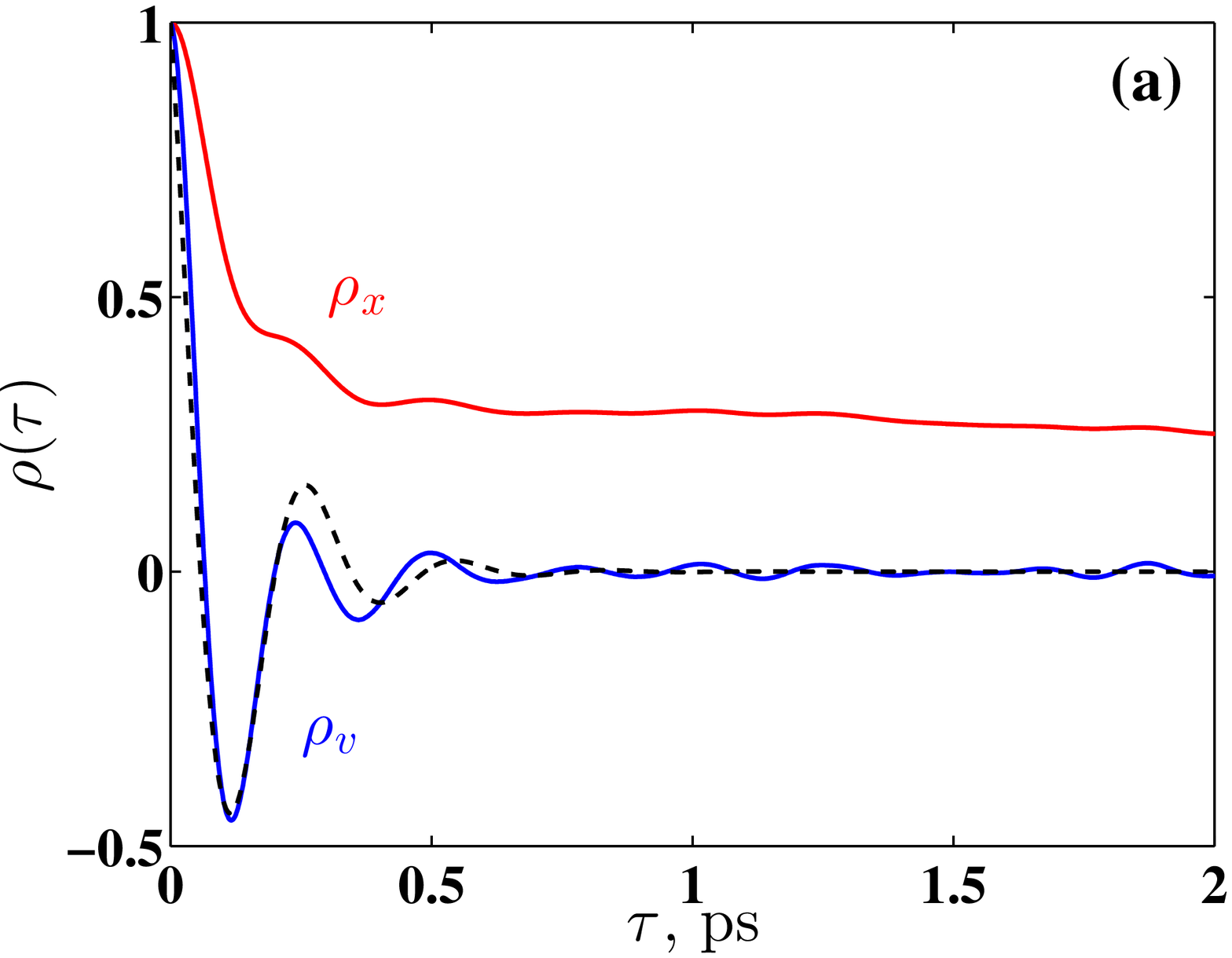}~~\includegraphics[width=6.5cm]{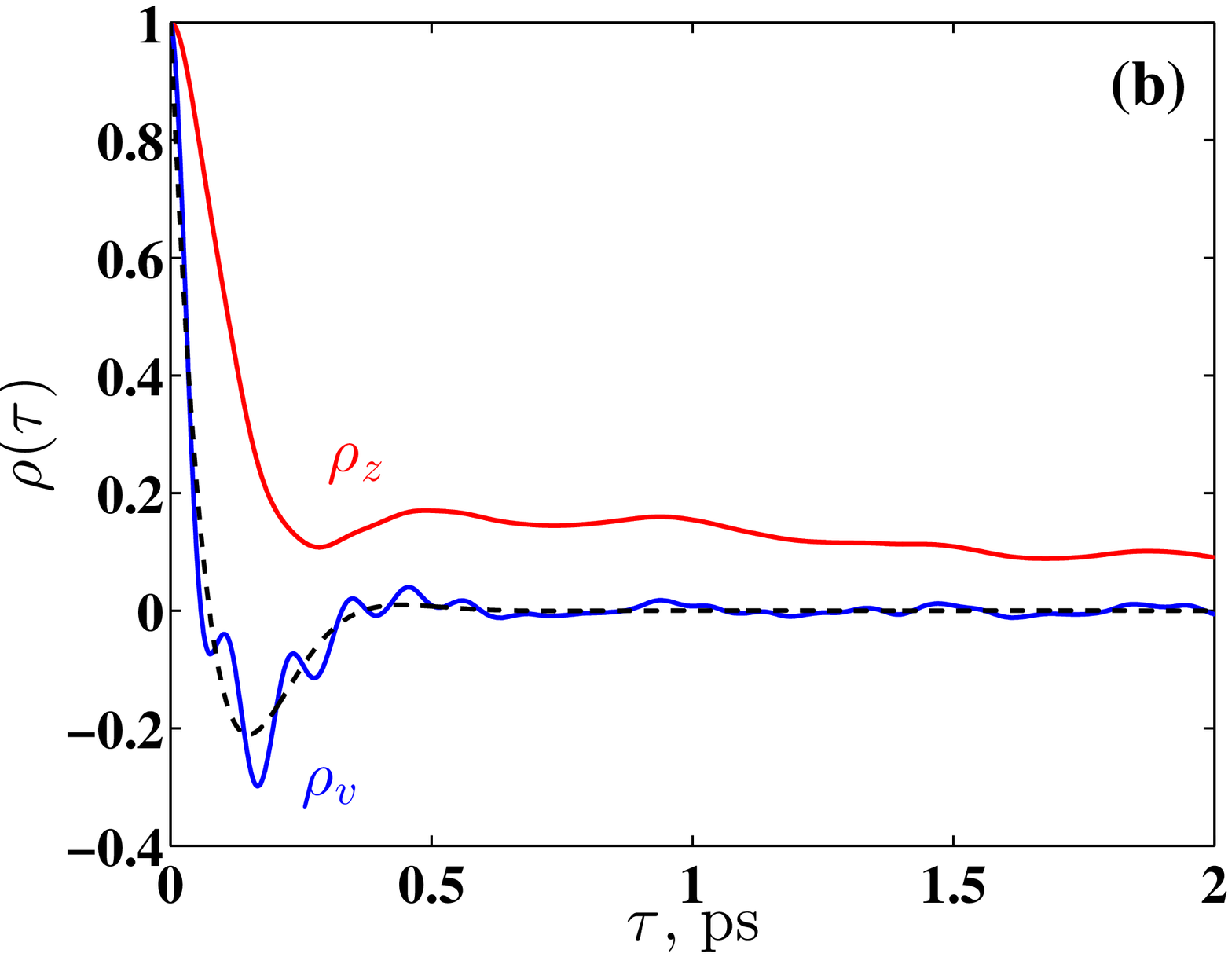}
\caption{(Colour online) The auto-correlation coefficient of: (a) the coordinate $x$ (red line) and velocity $p_x$ (blue line) of the ion in the site S1; and (b) the coordinate $z$ (red line) and velocity $p_z$ (blue line)  of the ion in the site S0.  Dashed lines are approximations of $\rho_v(\tau)$ by the expression (\ref{eq:acfv}). }
\label{fig4}       
\end{figure}

The power spectra shown in Fig.~\ref{fig3}(a) are typical of ions in each of the positions considered. The spectra for all coordinates have a part characterised by $1/f$ scaling, which for the velocity spectra transforms to $\propto f$ scaling, because $S_v(f)=S_r(f) f^2$. The existence of this scaling region is reflected in the behaviour of the auto-correlation coefficients $\rho(\tau)$ (Fig.~\ref{fig4}) for each coordinate, which decay slowly  towards zero. This picture  is typical of processes with long-range correlation. However, the auto-correlation for velocity does not support the presence of any long range correlation and $\rho(\tau)$ decays exponentially to zero, but in a oscillatory manner. This latter feature demonstrates that the motion of the atoms is not overdamped.

In general, the presence of the $1/f$ scaling would appear to suggest that the use of white Gaussian noise in the models (\ref{eq:gle}) and (\ref{eq:overd}) is questionable. In fact, $1/f$ part suggests~\cite{Kou:04} the use of Langevin equation with a fractional kernel in the memory function ${\bf M}(t-\tau)$ and corresponding fractional noise ${\bf R}(t)$.
However, $1/f$ scaling is observed for the low frequency range only. Further analysis has indicated that a possible origin of $1/f$ scaling lies in the relatively slow changes in the protein dynamics and, especially, in the selectivity filter. This can be illustrated by the spectral coherence function $\psi(f)$ \cite{Bendat_2011} (Fig.~\ref{fig3}(b)) which estimates the linear correlation between atoms for a given frequency. It can be seen that, for any atomic location, the coherence is maximal in the low frequency part of the coherence function $\psi(f)$. The $1/f$ scaling component can therefore be considered as a non-stationary component $f_n(t)$ in ion's dynamics and the PMF acting on the ion can be decomposed to non-stationary and stationary parts: $V({\bf r}_i)=f_n(t)+V_m({\bf r}_i)$. Removing the non-stationary component from the consideration allow us to use Langevin equation with time independent damping $\gamma$, white Gaussian noise $\bm{\xi}(t)$ and the potential $V_m({\bf r}_i)$ instead of a fractional noise and a fractional memory kernel for the potential $V({\bf r}_i)$.

The $1/f$ component has a large energy contribution that significantly increases the variance $\sigma_r^2$ of the coordinates. Hence this non-stationarity makes a big impact on the coordinates. However, the contribution of the $1/f$ component is negligible for the velocity time-series because the $1/f$ scaling of a coordinate converts to $\propto f$ for velocity spectra in the low frequency range (cf.\ spectra in Fig.~\ref{fig3}(a)). Thus, the use of velocity time-series for estimating parameters of a BD model  effectively removes the influence of the non-stationary component.

\begin{figure}[h!]
\mbox{}\hspace{-0.5cm} \includegraphics[height=4.5cm]{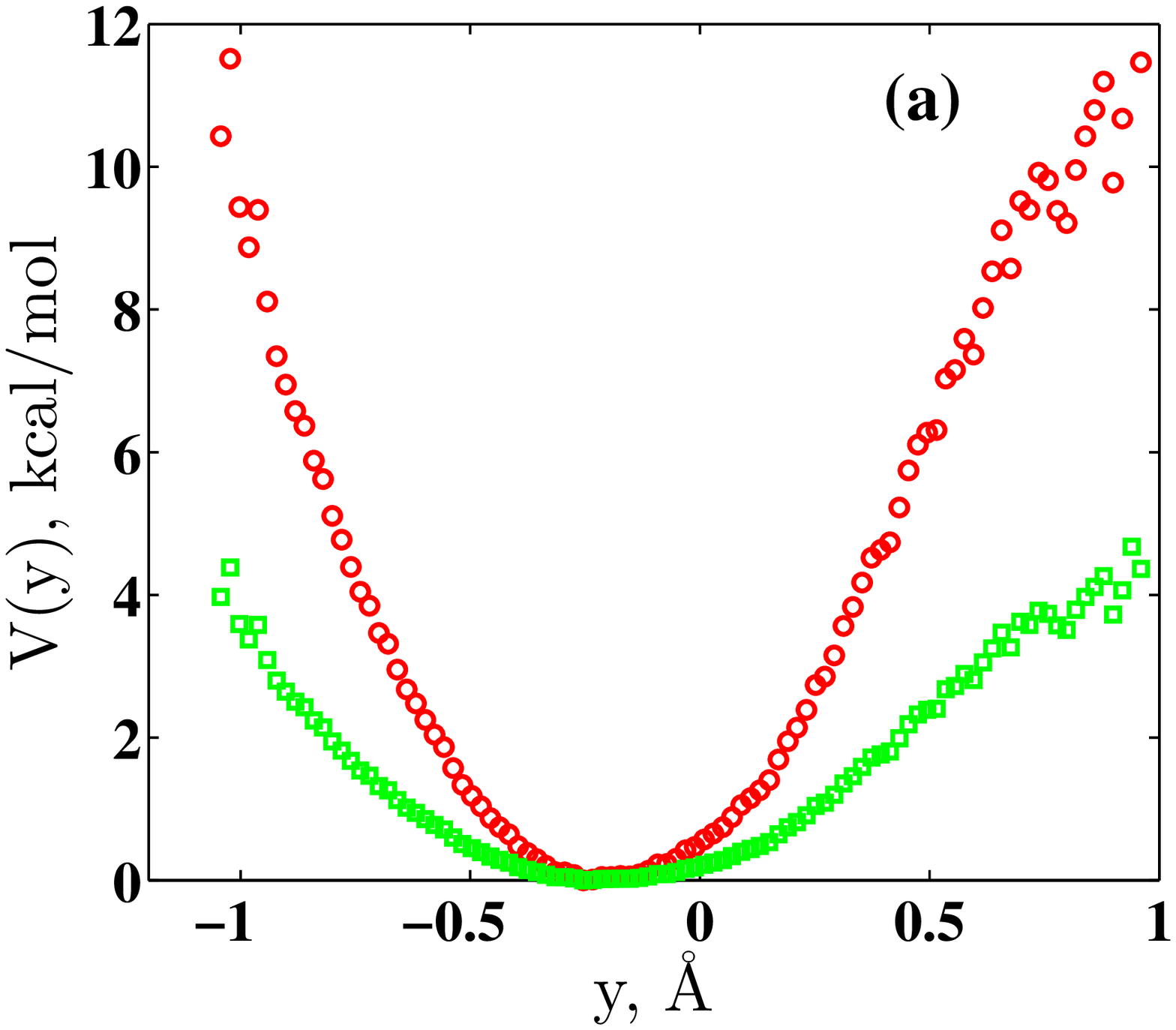}\includegraphics[height=4.5cm]{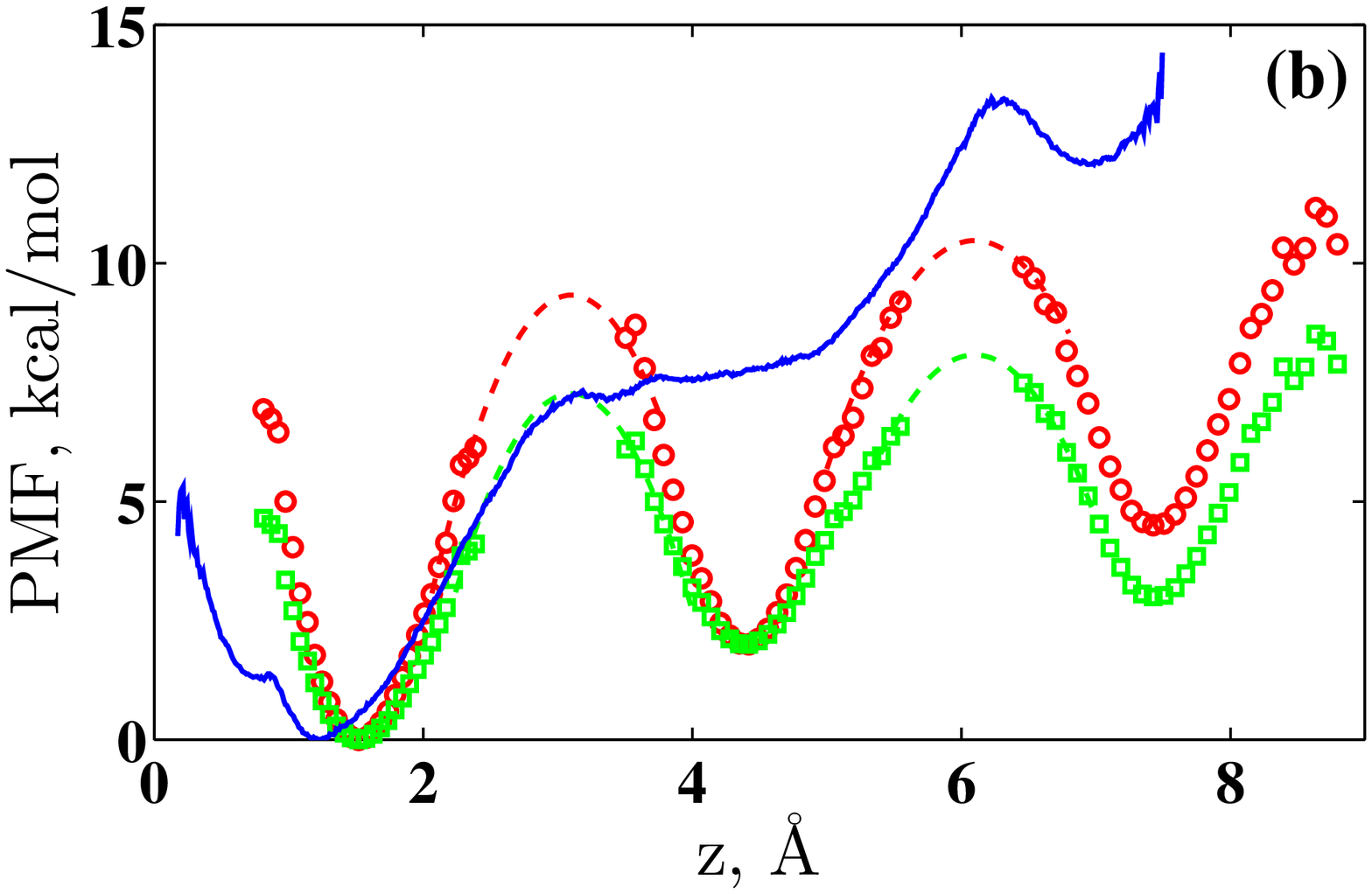}
\caption{(Colour online) The calculated PMFs  (a) $V_m(y)$ (red marker $\circ$)  and $V(y)$ (green marker $\square$) for the site S0. (b)  The potentials $V(z)$ (green marker $\square$), $V_m(z)$ (red marker $\circ$) and $V_{US}(z)$ (blue solid line) for an ion in the sites S2, S1 and S0 of the selectivity filter. The dashed lines in (b) correspond to approximations of the saddle states between sites. }
\label{fig5}       
\end{figure}

The plots of auto-correlation coefficients (Fig.~\ref{fig4}) clearly show that the overdamped Langevin equation (\ref{eq:overd}) only provides a very rough approximation, and that the motion of the ion within the selectivity filter should actually be described by an underdamped model of the following form:
\begin{eqnarray}
\label{eq:underd}
\dot{\bf r}={\bf v}_i, \ \ \ \
\dot{\bf v}_i = -\gamma({\bf r}_i) {\bf v}_i-\frac{1}{m_i }\frac{\partial V_m({\bf r}_i)}{\partial {\bf r}_i} + \sqrt{ \frac{2 k_BT\gamma({\bf r}_i)}{m_i}}\bm{\xi}(t).
\end{eqnarray}
For motion along the $x$ and $y$ axes, this model can be further simplified, because the distributions of $p(x)$ and $p(y)$ are Gaussian. Therefore the resultant model for ionic motion in the vicinity of a steady state inside the selectivity filter corresponds to a stochastic harmonic oscillator:
\begin{eqnarray}
\label{eq:sho}
\dot{ r}={ v_r}, \ \ \ \
\dot{ v_r} = -\gamma_r v_r -\frac{\Omega_r^2}{m} r + \sqrt{\frac{2  k_B T_r\gamma_r}{m} } \xi_r (t),
\end{eqnarray}
where $r$ denotes one of the coordinates $r \equiv x,y$, and $m$ is the mass of a K$^+$ ion. For motion along the $z$-axis, this model is approximate, since the distribution $p(z)$ is non-Gaussian but, because this deviation affects only the tails of the distribution, (\ref{eq:sho}) can be used for estimating $\gamma_r$. The model (\ref{eq:sho}) is well known as a system that produces so-called harmonic \cite{Schimansky_1990} or quasi-monochromatic \cite{Dykman_1991} noise, the properties of which are well-established including, in particular, expressions for the probability density $p(r,v_r)$, the variances of coordinate $\sigma_r^2$ and velocity $\sigma_v^2$, and the auto-correlation coefficient of velocity $\rho_v$:
\begin{eqnarray}
\label{eq:pdf}
p(r,v_r)&=&\frac{1}{\mathcal{N}_r} \exp\left( - \frac{\Omega_r^2 r^2}{2k_B T_r} \right) \exp\left( - \frac{m}{2k_B T_r} v_r^2 \right) , \\
\label{eq:variance}
\sigma_r^2&=& \frac{k_B T_r}{\Omega_r^2 }, \  \ \ \ \ \sigma_v^2=\frac{k_B T_r}{m} ,  \\
\label{eq:acfv}
\rho_v(\tau)&=& \exp \left( -\frac{\gamma_r}{2} \tau \right) \left[ \cos \omega_1 \tau - \frac{\gamma_r}{2\omega_1} \sin \omega_1 \tau \right], \ \ \omega_1=\sqrt{\Omega_r^2/m-\gamma_r^2/4}.
\end{eqnarray}
All of these expressions can be used for obtaining the parameters of the system (\ref{eq:sho}) for the $x$ and $y$ directions, and the system (\ref{eq:underd}) for the $z$ direction. The approach applied is straightforward. For any given direction, the first step is the estimation of temperature $T_r$ using the expression for the velocity variance $\sigma_v^2$ (\ref{eq:variance}). The second independent step consists of an estimation of $\rho_v(\tau)$ and then in fitting the calculated auto-correlation coefficient $\rho_v(\tau)$ by the expression (\ref{eq:acfv}); the ``nlinfit'' function of Matlab (2012a, The MathWorks, Natick, MA, USA) has been used (the fitted curves are shown in Fig.~\ref{fig4} by dashed lines). These two steps identify all of the parameters needed for BD modelling of the dynamics in the $x$ and $y$ directions. However, it is the motion along the $z$ coordinate that is the main subject of interest and, in this direction, the PMF $V(z)$ is non-parabolic (though usually close) as follows from the form of $p(z)$ in Fig.~\ref{fig2}(a). In this case, the expression (\ref{eq:PMFc}) could be used and in fact this expression is employed in all approaches for estimating the PMF, for example in the US method. We emphasise that the calculated spectra have a $1/f$ part for all coordinates, which implies that expression (\ref{eq:PMFc}) takes into account $1/f$ part for the PMF $V(r)$ by overestimating the variance $\sigma_r^2$. This is equivalent to overestimating either the temperature $T_r$ (if the value of $\Omega_r^2$ comes from $\rho_v(\tau)$) or the parameter $\Omega_r^2$ (if the value of $T_r$ comes from $\sigma_v$ or, as in the US method, is assumed equal to $T$, the temperature of the MD simulation). Therefore, the expression  (\ref{eq:PMFc}) requires a modification to remove the contribution of the $1/f$ spectral part. This can be effected by assuming that the distribution $p(r)$ is close to Gaussian, leading to a modified expression for the PMF:
\begin{eqnarray}
\label{eq:PMFcm}
V_m(r) = C_r - \sigma_r^2 \Omega_r^2 \ln[p(r)],
\end{eqnarray}
where $\sigma_r^2$ is calculated from the time-series of $r$ and $\Omega_r^2$ comes from fitting $\rho_v(\tau)$ by the expression (\ref{eq:acfv}). The PMFs $V(r)$, calculated by a {\em typical} technique via expression (\ref{eq:PMFc}), and $V_m(r)$ calculated  by the approach described above via expression (\ref{eq:PMFcm}), are compared in Fig.~\ref{fig5}(a). A visual comparison between $V(r)$ and $V_m(r)$ immediately reveals a large difference between them.

Because we consider the dynamics of steady states (with the ion centred in a site) only, there is no way to recover both $V(z)$ and $V_m(z)$ around saddle (boundary) states between the sites. We therefore roughly approximate $V(z)$ and $V_m(z)$ between the sites by the following procedure. First, the location $z_b$ of each boundary was calculated by estimating the average positions of oxygen atoms on the site boundaries. Secondly, a parabolic equation connecting the PMFs of one steady state with maximum at $z_c$ was calculated, and it was then used as an approximation of $V(z)$ or $V_m(z)$ between the sites. The resultant picture is shown in Fig.~\ref{fig5}(b) alongside with $V_{US}(z)$ as calculated by the US method (the solid line in Fig.~\ref{fig5}(b)). The difference between these two PMFs is highly significant.
Whereas the difference in the shapes of the PMFs for the saddle states can be accounted for on the basis of the approximations made in the estimation  $V(z)$ and $V_m(z)$, the differences around the steady states are too large to be explained by the assumptions made in estimating $V_m(z)$. In fact, there are virtually no errors for the PMF $V(z)$. The error of the PMF $V_m(z)$ comes from the non-Gaussianity of $p(z)$ (relative error 1--25\%) and the error in fitting $\rho_v(\tau)$ by (\ref{eq:acfv}) (relative error less than 10\%). Both errors are much smaller than the difference between $V_m(z)$ and $V_{US}(z)$.  Moreover, for $V_{US}(z)$, there is no steady state in the site S1, whereas the existence of this steady state follows directly from our MD simulations.

\section{Conclusions}
\label{concl}

Usually the analysis of MD simulations aims at the derivation of an order parameter, for example the PMF for an ion, whereas details of the ionic dynamics are not normally considered. We have shown, however, that a relatively simple analysis of time-series coming from unbiased MD simulations can be used to reveal important features of the dynamics. We can then take these features into account in a derivation of the PMF and corresponding BD model in the form of Langevin equation. We show that the dynamics of the atoms (ions and water molecules) in the selectivity filter includes a $1/f$ component, that, in principle, means  that a fractional memory kernel and a fractional noise should be used in Langevin equation. We argue that this $1/f$ component is unimportant for permeation events and, therefore, Langevin equation with a constant damping and a white Gaussian noise describes ion's dynamics.
The presence of the $1/f$ scaling leads, however, to a biased calculation of the PMF. We showed that use of the ion's velocity, rather than its coordinate, for data analysis allows us effectively to remove the $1/f$ component from consideration and, moreover, to obtain the parameters of a BD model. We also demonstrated that the conventional assumption of over-damped dynamics for the permeating ion is questionable and that an under-damped Langevin equation should probably be used instead. Our proposed approach to estimation of the PMF for an ion in the selectivity filter uses unbiased trajectories, and we were able to make a direct comparison of the resultant unbiased PMFs $V(z)$ and/or  $V_m(z)$ with the PMF $V_{US}(z)$ derived by use of the US method. We observed  significant differences between these PMFs.
Differences between unbiased PMFs and the biased US PMF cannot be explained by the assumptions which were put forward in the derivation of BD models. Further clarifications  of the observed differences are needed. Note, that the US method was based on parameters previously used in literature and it is possible that they are not optimal.

Finally, we comment that there are some important questions lying beyond the scope of this contribution. These are the interaction between the atoms (ions and water molecules) inside the selectivity filter, as well as activation events that might modify the PMF in the vicinities of the saddle states. Our preliminary  consideration of these  questions leads to the conclusion that the ion permeation should be described by the dynamics of coupled particles in a multistable potential.

The work has been supported by the Engineering and Physical Sciences Research Council (UK) under grant No.\ EP/G070660/1.
Computational facilities were provided by the MidPlus Regional Centre of Excellence for Computational Science, Engineering and Mathemtatics, under EPSRC grant No. \ EP/K000128/1.


\end{document}